\documentclass[12pt,preprint]{aastex}

\begin{document}

\title{Comparison of radiative accelerations obtained with atomic data
       from OP and OPAL.}

\author{Franck Delahaye $^{1,2}$ and Marc Pinsonneault $^1$ \\
 1 Department of Astronomy, The Ohio State University, Columbus OH 43210 USA \\
 2 LUTH, (UMR 8102 associ\'ee au CNRS et \`a l'Universit\'e Paris 7),
   Observatoire de Paris, F-92195 Meudon, France.}

\begin{abstract}

Microscopic diffusion processes (such as radiative levitation and
gravitational settling/thermal diffusion) in the outer layers of stars 
are important because they may give rise to surface  abundance anomalies.

Here, we compare radiative accelerations ($g_{rad}$) derived from the new
Opacity Project (OP) data \footnote{These data will be made generally 
available via a server or on CDROM. All request should be sent to Claude 
Zeippen (Claude.Zeippen@obspm.fr), or to Anil Pradhan 
(pradhan@astronomy.ohio-state.edu).} with those computed from OPAL and 
some previous data from OP. 

For the case where we have full data from OPAL (carbon, 5 points in the 
$\rho-T$ plane), the differences in the Rosseland mean opacities between OPAL
and the new OP data are within 12\% and less than 30\% between new OP and 
previous OP data (OP1) \footnote {The OP1 data are the 
data present in the database at Centre de Donn\'ees de Strasbourg (CDS, 
http://cdsweb.u-strasbg.fr/).}. 
The radiative accelerations $g_{rad}$ differ at up to the 17\% level when 
compared to OPAL and up to the 38\% level when compared to OP1. 

The comparison with OP1 on a larger $(\rho-T)$ space gives a difference of up 
to 40\% for $g_{rad}(C)$. And it increases for heavier elements. The 
differences increase for heavier elements reaching 60\% for Si and 65\% for S 
and Fe.

We also constructed four representative stellar models in order to compare 
the new OP accelerations with prior published results that used OPAL data.
The Rosseland means overall agree better than 10\% for all of our cases. 
For the accelerations, the comparisons with published values 
yield larger differences in general. 
The published OPAL accelerations for carbon are even larger relative to 
OP compare to what our direct comparisons would indicate. Potential reasons 
for this puzzling behavior are discussed.  

In light of the significant differences in the inferred acceleration rates, 
theoretical errors should be taken into account when comparing models with 
observations. The implications for stellar evolution are briefly discussed.
The sensitivity of $g_{rad}$ to the atomic physics may provide a useful test 
of different opacity sources. 

\end{abstract}


\section{Introduction}

Element segregation processes are clearly seen in stars and solar models
and occur from well understood physical processes.
Gravitational settling and thermal diffusion will tend to make heavier
species sink relative to the light ones.
Radiative pressure will tend to cause some species to rise in the stellar 
interior (we refer to this as radiative acceleration).
Radiative accelerations have been calculated using the Opacity Project 
(OP, Seaton et al. 1994, {\it{The Opacity Project Team}} 1995, 1997) data 
(Alecian \& Artru 1990; Alecian, Michaud \& Tully 1993; Alecian 1994; 
Gonzalez et al. 1995; Leblanc \& Michaud 1995; Hui Bon-Hoa et al. 1996; 
Seaton 1997,1999, Alecian \& Leblanc 2000; Leblanc \& Alecian 2004). 
Evolutionary calculations, however, have almost all been based on the OPAL 
theoretical opacities (Iglesias \& Rogers 1996) except for Seaton (1999).

In this paper we compare results from the up-dated data from opacity project 
(OP) with those obtained from OPAL and previous data from OP. We begin by 
discussing the astrophysical impact of element separation processes, and 
then move to our motivation for comparing with other datasets.

It is now generally accepted, as proposed by Michaud (1970) for Ap 
stars and Watson (1970) for AmFm stars, that radiative levitation plays 
an important role in hot and slow-rotating stars. The morphology of the 
Horizontal Branch of different globular clusters (GCs) presents some 
features that are not predicted by standard stellar models: 
Gaps in the blue tail (Ferraro et al. 1998), jumps in the
Str\"omgren color-magnitude diagram (Grundahl et al. 1999), surface
gravity anomalies (Moehler et al. 1995) and abundance
anomalies (Behr et al. 1999; Behr et al. 2000a,b).
This results in a bi-modal distribution in the HB stars of the studied GCs. 
 Qualitatively, 
 Hui-Bon-Hoa and co-authors (2000) showed that these observations could
 be the signatures of radiative acceleration ($g_{rad}$). 

Microscopic diffusion can also affect the internal structure of stars. The 
impact of gravitation settling on solar models has been extensively explored 
(Bahcall \& Pinsonneault 1992,1995; Vauclair 1998, Turck-Chi\'eze et al. 1998,
Sackmann \& Boothroyrd 2003). Diffusion deepens the solar surface convection 
zone, improving the agreement with helioseismic data on its depth, and yields 
a surface helium abundance in good agreement with the value deduced from
helioseismic studies. 
Element separation processes will affect the thermal structure, the 
convection depth, and the inferred initial abundances of solar models.
(Charpinet et al. 1997; Turcotte et al. 1998, hereafter T98a; 
Turcotte, Richer \& Michaud 1998, hereafter T98b;
Richer et al. 1998, hereafter R98). 

In addition, globular cluster age estimates are affected by the inclusion
of settling and diffusion (Chaboyer et al. 1992; Chaboyer, Sarajedini \& 
Demarque 1992; Vandenberg et al. 2002) leading to an age reduction of order 
10\% relative to models that neglect settling. Michaud et al. (2004) showed 
the importance of micro-diffusion in the age determination of open clusters 
and its effect on isochrone morphology. 
It has become clear that radiative levitation and diffusion processes must 
be included in stellar evolution codes in a self-consistent manner.

Computing these effects has been challenging; detailed comparisons
between theory and observations have had mixed success. 
T98b found that their predicted overabundances 
were larger than the observed ones. They explored other physical processes, 
such as turbulent diffusion and mass loss (T98b; Richer et al.
2000), as potential solutions. 
However, it is also possible that the uncertainties in the diffusion 
velocities themselves could be a significant error source. 
Because the two effects (gravitational settling and levitation)
are in opposite senses, a small difference in the radiative 
accelerations could change the magnitude (or even the sign) of the predicted
abundance anomalies by a much larger than linear factor when the two effects 
are of similar magnitude.
Before any attempt to generate a model with micro-diffusion including 
radiative levitation, a careful study of the atomic data available is 
necessary. 
We therefore evaluate the uncertainties in the radiative acceleration 
coefficients themselves. 

In section 2 we describe the method used to obtain the radiative accelerations 
and the different stellar model structures. In the 
first part of section 3 we compare and analyze the results from the new data
and from the previous OP data. In a second part of this section we compare 
the results from OPAL and NEW for carbon. This is the only element for which 
we have some monochromatic opacities from OPAL. For the other 
elements we compare our results with published works in the third part of this 
section. 
The discussion and the conclusion constitute sections 4 \& 5 respectively.

\section{Method.}

In a star, the net radiative flux is going outward. As the photons move 
toward the surface, they interact with the ions present in the star. 
Each species, depending on its state of ionization, the excitation level and 
its corresponding absorption cross section, experiences a net force that tend 
to make ions rise when they absorb momentum from photons.
The resulting acceleration is called the radiative acceleration and is defined
as follows (Seaton 1997):

\begin{equation}
 g_{rad}(k) = \frac{F}{c}\frac{M}{M(k)} \kappa_R \gamma(k)
\end{equation}

where M is the mean mass per atom, M(k) is the atomic mass of atom $k$, 
$\kappa_R$ is the Rosseland mean, F is the total flux of the radiative source 
(a blackbody at $T=T_{eff}$) and $\frac{F}{c}$ corresponds to the total 
momentum radiative flux associated to it. Finally, $\gamma(k)$ is a 
dimensionless quantity characterizing the individual contribution of element 
$k$ to the total opacity defined as follow (Seaton 1997):

\begin{equation}
\gamma(k)=\int \frac{\kappa _{\nu}(k)}{\kappa _{\nu}(total)} f_{\nu}d\nu
\end{equation}
\begin{equation}
\kappa _{\nu}(total)=\sum _k \kappa _\nu (k) + \kappa_{scat}
\end{equation}

where $\kappa _{\nu}(k)$ is the monochromatic opacity of element $k$, 
$\kappa _{\nu}(total)$ is the total opacity, both in $cm^2/g$ of mixture, 
and $f_{\nu}$ is a weighting function defined below. As a reminder, with 
this notation, $\kappa_R$ is defined as:

\begin{equation}
\frac{1}{\kappa_R}=\int \frac{1}{\kappa _{\nu}(total)} f_{\nu}d\nu
\end{equation}

\begin{equation}
f_{\nu}=\frac{15 h^5 \nu^4}{4\pi^4 k_{B}^5 T^5} \frac{e^{h\nu/k_BT}}{(e^{h\nu/k_B} - 1)^2}
\end{equation}
In the present work, we compute $g_{rad}$ with and without including 
the effect of momentum tranfer to electrons during photoionization processes. 
In other words, we have use $\kappa _{\nu}(k)$ or $\kappa^{mta} _{\nu}(k)$
(as define in Seaton 1997, equation(31)). This allow us to isolate each 
difference in the atomic data. For the OPAL data as for the OP data we
subtracted the electron scattering opacities and the opacities corresponding 
to the momentum transfered to the electrons during the photoionization from
the monochromatic opacities of the studied elements. However This is not 
subtracted from the total opacity. We used exclusively the OP data to remove 
these contribution. We did not remove the OPAL electron scattering opacity from
the OPAL data because it is already subtracted via the OP data.

In order to estimate the systematic errors in the radiative accelerations
due to atomic data, we compare the values of $g_{rad}$ obtained with four sets
of atomic data, two from OPAL and two from OP (NEW and OP1). 
We could compare with OPAL data for carbon (Iglesias 2004, private 
communication) at the five points in the ($\rho$-T) plane listed in Table 1,
and discuss indirect comparisons with the literature in \S 3. Ideally, one 
would directly compare the monochromatic opacities (total and specific for 
each element) from OP and OPAL. However, due to technical problems the OPAL 
group could not provide a full set of data.

We first calculated the Rosseland mean opacities and $\gamma(C)$ with the
3 sets of data. We then calculated the accelerations for different sets of 
$T_{eff}$ and radius (these determine the flux in eq.1). 

For the other elements present in the mixture, we do not have the relevant OPAL
atomic data. In section 3.2 we thus compare OP data for other elements with
previously published OPAL results (T98a; T98b and R98).
We calculated the Roseland mean opacities, $\gamma(k)$ and the
radiative accelerations using OP data for different types of stars. 
This indirect comparison allows us to span a large portion of the ($\rho$-T)
plane. As we will show, however, it is difficult to directly interpret 
differences obtained in this manner.

We calculate the structure of different stars with the YREC code (see
Bahcall, Pinsonneault \& Basu 2001 for a description of the model ingredients) 
in which we modified the micro-diffusion subroutines. 
We extended the composition vector to include all
species present in the mixture in order to track individually their evolution
within the model as a function of time. The diffusion equation for element 
separation processes has been modified in order to treat separately each 
species of the initial mixture. Instead of treating the gravitational 
accelerations of all element as if they were settling like fully ionized Fe,
the gravitational settling coefficient are calculated individually for each 
element. We also included the individual radiative accelerations 
calculated using the OP data and the method described by Seaton (1997, 1999). 
All the technical details of the calculation will be presented in a subsequent 
paper.

We first calibrated our model with a mass of 1.0 $M_{\odot}$ to reproduce the 
observed solar luminosity ($L_{\odot}$), radius ($R_{\odot}$) and 
surface $Z/X$ ratio at the solar age  ($4.57\ Gyr$). The
initial composition used here includes 17 elements
(H, He, C, N, O, Ne, Na, Mg, Al, Si, S, Ar, Ca, Cr, Mn, Fe, Ni) with relative
abundances from the
Grevesse \& Noel (1993) mixture. The calibrated model yields the initial 
helium mass fraction ($Y_\odot=0.271$) and the mixing length ($\alpha_0=2.09$).
We used this solar calibration for our other models.
As previously noted (T98a), it might not be justified to use 
these parameters from a calibrated Sun for stars with a range of masses and 
evolutionary states. But by doing so we transform the absolute errors in 
the input physics into relative ones, which is a significant advantage for a 
solar calibration. For all models as well as for the 5 special points for 
which we have the monochromatic opacities from OPAL, we used the composition 
given in Table 2.

In the present work we focus primarily on quantifying the errors in the 
coefficients themselves, so we have not yet included the feedback 
on the opacity induced by changes in the heavy element mixture. 
Changes in the relative heavy element abundances can certainly be 
important in some contexts as suggested by Alecian, Michaud \& Tully (1993). 
For example, 
Richer, Michaud \& Turcotte (2000) found an iron convection zone in AmFm models
including micro-diffusion that was caused by feedback effects.
We intend to include such effects in future work focused on the applications 
of diffusion to stellar evolution problems.

\section{Results}

Our aim in this section is twofold: we present the effect of the 
up-dated OP opacities on the radiative acceleration and then compare the new 
results to OPAL. We present two specific cases, C and Fe, to show how the
monochromatic opacities modify the radiative acceleration.
The improvement in the OP atomic data due to the inclusion of inner-shell
transitions systematically enhance the radiative accelerations relative to 
the older OP data. 
The impact of the new physics increases for heavier species and higher
temperature. 
When directly compared to the OPAL data, the accelerations for carbon are
in reasonable agreement.  However this agreement deteriorate in the comparison 
with published work. The agreement between OP and OPAL is expected to be, and 
is, less favorable for heavier elements.

\subsection{Comparison between New OP and OP1.}

In order to understand the contribution of different ingredients to $g_{rad}$,
we present the monochromatic opacities for C and Fe at a specific ($\rho-T$) 
point.
The main difference between the two sets of data (OP1 and New OP) is the 
inclusion of the inner-shell transitions in the later. 
  
For our purpose, it is worth recalling that the acceleration 
$g_{rad}\propto \kappa_R \times \gamma$ and 
$\gamma \propto \int \frac{\kappa_{element}}{\kappa_{tot}}f_{\nu}$.
The interplay between the various terms in equation (1) is illustrated in 
Figure 1. The data used for this figure are from the new OP data and
OP data without inner-shell transitions (Seaton, private comunication).
The top panels show the monochromatic opacities for C (left) and Fe (right). 
The middle compare
the total monochromatic opacities (same on each side) and the bottom panels 
illustrate the ratio of the 2 which corresponds to part of the integrand in 
the definition of $\gamma$ (see equation (2)) as a function of u ($u=h\nu/kT$).
The physical conditions ($log(T)=6.3$ and $log(R)=-1.5$) correspond to point 5
in Table 1 and are close to the conditions at the base of the convection zone
of the Sun.
These monochromatic
opacities have been obtained by interpolating OP mesh data (OP5 data, Seaton
private comunication) in Ne 
(electron density corresponding to $log R = -1.5$). Using these interpolated
opacities to calculate $\kappa_R$, $\gamma$ and $g_{rad}$ generate an error 
smaller than 2\%. We estimated this error by comparing $\kappa_R$, $\gamma$ 
and $g_{rad}$ obtained with the interpolated opacities to the values derived
from the interpolation of 4 values of $\kappa_R$ (or $\gamma$ or $g_{rad}$)  
calculated on the grid point. 

From the top and middle panels, one can see that new OP values
are higher than OP1 for both the individual and total monochromatic 
opacities. However increase in the specific elemental opacities is small 
for C, large for Fe and intermediate for the total opacities. This arises 
naturally from the changes in atomic physics .The inner-shell transitions 
included in the new OP data enhanced significantly the Fe monochromatic 
opacities, but lead to a small effect for C as expected (C is almost 
fully ionized). 
By definition, $\gamma$ is governed be the ratio of the individual 
to the total monochromatic opacities weighted by $f_{\nu}$. The weighting 
function $f_{\nu}$ decreases rapidly at low and high frequency, damping all 
differences for these regions.

At lower temperature, the importance of inner-shell decreases and for the 
physical conditions typical of envelopes the two sets of data are in good 
agreement (within 20\%). The previous data were meant to be used for this 
purpose (Seaton et al. 1994).

In the case of carbon the differences between the 2 ratios (NEW and OP1) will 
be reduced when we compare the $\gamma$ factors.
This can be seen in the bottom left panel: 
$\kappa_{\nu}^{OP} \leq \kappa_{\nu}^{NEW}$ and 
$\kappa_{tot}^{OP} \ll \kappa_{tot}^{NEW}$ but
$\frac{\kappa_{tot}}{\kappa_{tot}}^{OP} <\frac{\kappa_{tot}}{\kappa_{tot}}^{NEW}$

For carbon, the increases in the total monochromatic opacities reduces 
significantly.
As a consequence, the difference in 
$g_{rad}\varpropto \kappa_R \times \gamma$ will be smaller than in $\kappa_R$.
The changes in $\gamma$ are partially compensated by those in $\kappa_R$

In the case of iron, the fractional changes in the Fe monochromatic opacities  
are greater than those for $\kappa_{tot}$, which makes $\gamma$ larger. 
This makes $g_{rad}$ significantly larger compared to the OP1 values.
 
We calculated the acceleration for 15 elements present in the mixture for a 
range of ($\rho-T$) characteristic of the physical conditions found in 
different stellar models.
The details of theses models are given later in this section.
For all the different structures, $\kappa_R$ differs on average by less
than 30\% (see Table 3) with particular points differing by up to 38\%
(see Figure 1). For the acceleration, the results depend on the element and
we can divide them in 2 groups.

\subsubsection{Light elements: C to Al}

The lighter metals follow the pattern seen in the carbon data. The differences 
in the acceleration are less than 40\% with a rms smaller than 40\% in a sense
that the new data are larger. 
These differences are dominated by the
Rosseland mean and consequently the global effect is not. The monochromatic 
opacities for these elements are not significantly modified by the 
inclusion of the inner-shell transitions. The
total monochromatic opacities are affected by the other abundant species
for which the inner-shell transitions are important. As a consequence, the 
ratio of the two presents a difference that is balanced by the differences 
in the Rosseland mean opacity.
The final results are closer than the Rosseland mean opacities themselves.

\subsubsection{Heavy Species: Si to Ni}

For the heavy elements, the accelerations increase
by up to 80\% with an rms between 10\% to 65\%. These elements have 2 regimes.
At lower temperatures, the increase in the monochromatic opacities is of the
same order than in the total monochromatic opacities and the ratio of the two,
which constitute the integrand in $\gamma$, stays relatively constant. Then the
differences in the accelerations follow the trend of $\kappa_R$ which does not
differ a lot between the 2 datasets.
At higher temperatures, where the contribution of the inner-shell transitions
are very important, the ratio is not similar and $\gamma$ differs 
significantly
($\gamma_{NEW} > \gamma_{OP1}$). The increase in the elemental monochromatic
opacities is much more important than the rise in the total monochromatic 
opacities. $\kappa_R$ is also larger, resulting a significant 
increase in the acceleration.

\subsection{OP and OPAL: Direct comparison for C.} 

The monochromatic opacities for carbon obtained from the OPAL group allow 
us to directly gauge the effect of the difference in atomic data in calculating
the radiative accelerations. 
 
The OPAL data have been re-sampled and interpolated in order to match the 
frequency points used in the OP data. This does not introduce any significant 
error (less than 2\% in all parameter). We estimated this error by recomputing
the Rosseland mean opacity and comparing it to the value provided with the 
data by the OPAL group. In Table 4, the values in parenthesis are the 
recalculated values. 
We also reproduced Fig. 3 
form Iglesias \& Rogers (1995). We are confident that any error introduced
by this procedure are negligible and allow a fair comparison.

In Table 4 we compare the OP and OPAL Rosseland mean opacities values.
The results for the Rosseland mean show a difference that does not exceed
12\%, in the 
sense that the new OP opacities are lower than OPAL (the differences 
between NEW and OP1 range from 6\% to 30\%). A detailed discusion on the 
comparison of Rosseland mean can be found in Badnell et al. (2004).
In particular, the difference for the solar case is only 2.5\%

The results for $\gamma$ and $g_{rad}$ are presented in Tables 5 to 8.
The $\gamma(C)$ factors differ by less than 12\% when compared to OPAL and
by less than 15\% for OP1. The momentum transfert to the electrons (mte) does
not change the results.
The net results on the accelerations range from 5\% to 12\% when compared to 
OPAL without the momentum transfer to the electrons (mte) correction and from 
2\% to 17\% when this effect is included.
When compared to OP1 the differences are less than 38\% with mte corrections. 

\subsection{OP vs OPAL: Other elements.}

The comparison with OPAL for other element can only be infer
indirectly at the moment. The OPAL atomic data are not available. But from
Seaton and Badnell (2004), their figures 5,6,7,10 gives an insight of the 
expected differences. For these simple mixtures, one probes the difference of 
the element contribution in either dataset. As we have shown above, the 
difference on the carbon monochromatic opacities are small (Seaton and 
Badnell (2004) their figure 5). The same results is expected for O 
(Seaton and Badnell (2004) Fig. 6).
The differences in S increase ( Seaton and Badnell (2004) Fig. 7) and become 
significant for Fe (Seaton and Badnell (2004) Fig. 10).
All these results are for the condition of the 5 points from Table 1.
An interesting point is that the new data produces larger
monochromatic opacities for C, O and S compared to OPAL but smaller
values for Fe. 
To extend the comparison, we have calculated the accelerations for 
different element for 4 stellar models and compare them to previous published
works.  

\subsubsection{Models.}
For the other elements the four types of models used here are: 

1) The Sun at 4.57 Gyr;

2) A model (R98) with $T_{eff}=10000K$ and $log(R)=-3$ where R
is defined (in previous OPAL works) as $R=\frac{\rho}{T_6^3}$ with  
$T_6=\frac{T}{10^6}$ where $\rho$ is the mass density in $g\ cm^{-3}$ and 
T the temperature in Kelvin.

3) A $M=1.3M_{\odot}$ model at 70 Myr ($T_{eff}=6500K$);

4) A $M=1.5M_{\odot}$ model at 30 Myr ($T_{eff}=7080K$);

We picked regimes where the diffusion effects are known to be important.
While levitation is a small effect in the Sun, we have precise data to compare 
theory and observations. The solar models therefore provide a useful point of 
comparison with other investigators.
The other models are in regimes where the radiative levitation is most likely 
to play an important role; the different cases span 
a large domain of the ($\rho,T$) space. 
The second case is designed to mimic the physical conditions appropriate for
hot horizontal branch stars or intermediate main-sequence stars of mass around
2.5 $M\odot$.
The third and fourth cases are 
models of typical F stars where levitation is producing overabundances of 
Fe-peak elements for the slow-rotating Fm stars (T98a). 
 
The results from T98a,T98b 
and R98 are taken from the articles directly, using a digitalization of 
Figures 11 \& 12 from T98a, Figure 1 from T98b and Figures 1 \& 7 from R98 
( the original data for the structure, acceleration and monochromatic
opacities were not available to us). 

The results for the Rosseland mean opacities are presented in Fig. 1. There
are 4 panels representing the percentage difference in $\kappa_R$ 
of the 4 stellar models studied.

\subsubsection{Results.}
The OPAL $\kappa_R$ have been calculated using the tables and the routines
provided on the OPAL website \footnote{http://www-phys.llnl.gov/Research/OPAL/}
. The results for the accelerations for the solar model at 4.57 Gyr are 
presented in Fig. 3. We have plotted 
the ratios $\frac{g_{rad}}{g}$ using OP and OPAL data for several elements, 
where $g$ is the gravity.
There is a maximum difference of 10\% in $\kappa_R$,
but the difference in $g_{rad}$ rises to 60\% for Fe and up to 200\% for 
C between the accelerations calculated with the new OP data and the data 
extracted from T98a, in the sense that the OP values are smaller. For physical 
conditions appropriate for the base of the solar convection zone, the 
difference (defined as $\frac{NEW-OPAL}{NEW}$) is 20\% for Fe and 150\% for C.

The accelerations for the second model with
$T_{eff} = 10000\ K\ Log(R)=-3$ are presented in Fig. 4.
The Rosseland means agree within 15\% and the radiative
accelerations within 50\%.

Figures 5 \& 6 give the results for the next two models ($M=1.3\ M_{\odot}$ 
and $M=1.5\ M_{\odot}$). The difference in $\kappa_R$ (Fig. 2) reaches a 
maximum of 35\% while the accelerations differ by less than 70\%. 

For all cases, regardless of the type of star or the region of the ($\rho,T$)
plane explored, the differences in the Rosseland mean opacities between OP and
OPAL data are much smaller than the disagreements between the corresponding 
accelerations. As described in section 3.1.1, $\kappa_R$ cannot alone be held 
responsible for the large gap between $g_{rad}^{OP}\ and\ g_{rad}^{OPAL}$. 
The combination of the discrepancies in $\kappa_R$ and $\gamma$  
defines the differences in the radiative acceleration. 
The accelerations in the different stellar model show large discrepancies for 
most elements. The largest ones are found in the solar model. 

\subsubsection{General trends.}

In section 3.1 we anticipated the differences in the acceleration for the 
elements heavier than carbon to be larger than for carbon itself.
For the last three cases ($T=10000K$, $M=1.3M_{\odot}$ and $M=1.5M_{\odot}$) 
the results seem to follow roughly the trend (see Fig. 4,5 and 6). However the 
differences are well within 50\% with a rms of the difference smaller than
30\%.

However for the solar case the differences in carbon accelerations are the 
largest. The difference for carbon range between few \% to 250\% through the 
solar structure while all of the other 
elements have differences within 60\% (the regions near the center, small 
$r/R_*$, are uncertain due to the difficulties to digitize this region).
It is difficult to understand these results. 
Our direct comparison for carbon yielded a difference of 2\% for case 5 where 
the the physical conditions are similar to those of the base of the convection 
zone of the solar model.
The results from T98a give a difference of 150\% at similar physical 
conditions.
In other words, our accelerations using OPAL data are smaller than those 
presented by T98a for similar conditions. The points
are not exactly at the same $\rho-T$, but it seems unlikely that this can 
explain such a large difference because the changes of T and $\rho$ are very 
small. We will address this issue in the next section. 

\section{Discussion.}

\subsection{Stellar Model differences.}

The first source of differences we can think of is the stellar code itself. 
We digitized figures from the published papers and attempted to 
construct equivalent physical cases for our models.
Our models were derived from a different stellar evolution code than
the one used by T98a, T98b and R98.
In principle, differences in opacities
and accelerations could therefore simply arise from differences in the 
structure of the models themselves ($\rho,\ T,\ X_i$). The total flux
could also change the absolute scaling of the accelerations. The digitization
process introduces errors. Even if the OPAL data were provided by the same 
authors (Carlos Iglesias), it could be possible that our OPAL data differ from 
theirs.
Finally it is also possible that there are genuine differences in the 
calculated $g_{rad}$ that arise from errors in either calculation.

Modern solar models have very similar thermal
structures for similar input physics, so we believe that structural
differences are not an explanation for the differences in
the solar case. When comparing our models with T98a (model H, Table 6), the 
base of the convection zone is at similar
depth ($\frac{r}{R_{\odot}(Model\ H,\ T98a)}=0.7176$,
$\frac{r}{R_{\odot}(present)}=0.7146$), the temperature are in very good
agreement ($log(T_{CZ}^{T98a})=6.3343$, $log(T_{CZ}^{T98a})=6.3360$),
the central density and temperature are within 1\% and the composition is
within few percent at the surface.
We also checked that a 10\% difference in the composition does not produce 
any significant difference in the acceleration.
The constraints on solar models are the strongest compare to other stellar 
models.
The two models use similar input physics (The Krishna-Swami $T-\tau$ relation 
is used for the atmosphere, the energy generation routines are the same and 
both use the mixing length theory). 
The equations of state differ but the effect is expected to be, and is, 
very small (for the center and the surface convection zone, $\rho$ and T
are within 1\% in both models). 
Indeed for the interior solar conditions, the gas can be assimilated to a
fully ionized ideal gas. 
It is very unlikely to have large difference in the structure.
We calculated the partial derivatives of the acceleration at few points to 
gauge the uncertainties due to the structure. The partial derivatives are 
calculated as simple differences 
$\frac{\delta log(g_{rad})}{\delta log(X_i)} = \frac{\Delta log(g_{rad})}
{\Delta log(X_i)}$ where $X_i$ is T or $\rho$ or $r/R_*$. Two of the variables
are kept constant while calculating the partial derivative.

At the base of the surface  CZ of our solar model we have : 
$\frac{\delta log(g_{rad})}{\delta log(T)}\vert _{\rho,r} =-5 $,
$\frac{\delta log(g_{rad})}{\delta log(\rho)}\vert _{T,r} =0.5 $ and 
$\frac{\delta log(g_{rad})}{\delta (r/R_*)}\vert _{\rho,T} =-1.22$.

This implies that a deviation of 20\% in the temperature or a factor of 7 in
density or a difference of 112\% in the radius or flux is needed in the model 
in order to invoke the structure as the source of differences in the radiative
acceleration. As we mentioned above the differences are less than 1\% for 
the temperature, 2\% for the central density and less than 1\% for the position
of the base of the convection zone.

For the second model the structure is just a vector defined identically in
our model and the published one. It consists of a set of temperature points
($log(T)=$4.3 to 7.3) at a constant R ($log(R)=-3$, $R=\frac{\rho}{T^3_6}$, 
$T_6=\frac{T}{10^6}$).

For the 2 other stars, the final model $T_{eff}$ are identical to those from
T98b for a similar age. So we expect the structures to be very
similar; the effective temperature sets the convection zone depth, so the
thermal structures should be similar at a fixed $T_{eff}$.

Because the structure for the 2 last models are not as well constrained as the 
solar model, and because the differences in the acceleration are not too large
it is a potential source of the differences between our calculation and the 
results from the literature. 
However, for the solar case, it is ruled out.

The composition is also not the right candidate because they are equivalent 
to a solar composition in our model and in the 3 other models (2 last models 
are taken at the zero age main-sequence and by definition for case 2).
For the solar case the composition does not change by more than 10\% and we 
checked that this does not affect the acceleration.

The errors due to the digitization are at the level of 3\%, which is 
measurable but much less than the differences that we are seeing. 

The OPAL monochromatic opacities may be different from those used by the 
previous investigators. The data we received (Iglesias 2004, private 
communication) were re-created for 5 specific $\rho-T$ conditions (Table 1). 
Despite the fact that we were able to reproduce Iglesias \& Forest Figure 3a 
(Iglesias \& Forest 1995) it is possible that the OPAL data used by Turcotte 
and others differ from ours. 
In the second case, where the structures are identical by definition,
we found differences in the OPAL Rosseland mean opacities.
The extracted value from the plot (R98 Figure 1) is 
$\kappa_R=3.45$ at $log(T)=6.0$ and $log(R)=-3$ while our value is 
$\kappa_R=3.72$. 

Finally, we should not exclude the possibility that the two results are
really different for the same physical conditions which could indicate an
error in one or both calculations.

We have been through the different possible sources of errors and we will not
speculate further on these results given the number of inconsistencies.
A direct comparison with the 2 sets of data would be more reliable
and we are expecting to be able to run it for other elements than carbon and
for a large portion of the $\rho-T$ plane. We will be glad to provide our data
to allow others to run parallel comparisons.

\subsection{Further discussion.}
 
The frequency resolution used to sample
the monochromatic opacities, as already mentioned by Seaton (1997,1999), is
another important source of possible difference in the accelerations.
The calculations of $\kappa_R$ and $\gamma$ use a sampling technique
(Seaton 1997; Leblanc, Michaud \& Richer 2000) .
All the cross sections are sampled at a specific frequency mesh and then
integrated in order to calculate the Rosseland mean and $\gamma$. While a
modest frequency resolution might be sufficient for the determination of
$\kappa_R$, higher resolution is required for the acceleration.
Indeed, as discussed in Seaton (1997), $\gamma$ is much more sensitive to the
frequency mesh than the Rosseland mean. The specific opacity will
use a smaller number of line than $\kappa_R$. If the frequency mesh is bigger
than some line width, the sampling might miss some lines. On average, it will
have a small impact on the total opacity given the large number of lines
already included. However it will be more crucial for the specific opacity
which has less lines than the total opacity.

While both OP and OPAL agreed on the convergence of $\kappa_R$ within
2\% (Rogers \& Iglesias 1992; Seaton et al. 1994) using $10^4$ points over
the range of significant frequency, using an constant mesh (u), Seaton (1997) 
compared the effect of varying the frequency resolution from $R=10^4$ to 
$R=10^5$ to $R=10^6$ on $\gamma$.
He showed that it could lead to a factor of few difference in $\gamma$, and
varies depending on the element and the physical conditions. Seaton also 
addressed this issue in appendix B of his 1999 article. 
While the frequency resolution used in OPAL data (constant u mesh with $10^4$ 
points over the all frequency range) is adequate to determine with good 
accuracy the Rosseland mean opacity, it certainly will limit the accuracy of 
the acceleration when the composition changes during the evolution 
calculations.
The new OP data use an equally spaced mesh (v) which samples preferentially 
the region where $f_{\nu}$ is not too small (see Badnell et al. 2004). This 
increase significantly the accuracy of $\gamma$ without requiring a prohibitive
number of points. We have tested the results of the 
accelerations using $10^4$ points in v-mesh and compared with the acceleration
using $10^5$ points in u-mesh. The results are in good agreement.    
We then expect more accurate accelerations derived from OP data than from OPAL.

\section{Conclusion.}

The phenomenon of radiative levitation can induce significant changes in the
surface abundances of stars, and the underlying theory that predicts 
levitation of some elements is based upon well-understood physics. However, 
quantitative theoretical predictions rely upon subtle details in the radiative 
opacities, and it has been difficult to estimate the error induced by 
uncertainties in the atomic physics and equation of state. The availability of 
two independent theoretical sources for opacity data (OP and OPAL) therefore 
provides an important test of systematic errors and their potential impact on 
this problem, as well as other aspects of the theory of stellar structure and 
evolution.

We have calculated the radiative accelerations using the up-dated OP 
opacities which include all contributing inner-shell processes. The impact
on $g_{rad}$ is important and generates an increase by up to 80\% compared
to the data without the inner-shell. The light element are less affected than 
heavy species. The former changes by up to 40\% while the later are affected to
a 80\% level.  

It is important to note that the difference in accelerations was significantly 
larger than the difference in the Rosseland mean opacities; we have traced 
through the interplay of factors that is responsible for the increased 
sensitivity of radiative acceleration computations to changes in the underlying
atomic physics relative to the mean opacity.

In the case of carbon, we were able to directly compare the OP and OPAL data
for some cases.  We found that the accelerations derived from the new OP data 
differ by less than 17 \% compare to the accelerations derived from 
OPAL data.  As expected, we found larger differences for heavier species in 
indirect comparisons with published literature values, ranging from 10\% to
70\% than the OPAL ones, depending upon the element and the physical 
conditions. 

However, these indirect comparisons also gave larger differences for C than 
the ones that we computed directly. In other words, our reconstruction of the C
accelerations from the OPAL data differ from the published accelerations 
calculated using OPAL data in the sense that our OPAL accelerations are 
smaller. For the case that can be compared most directly (our solar case), the 
differences were 2\% while the differences with the published work reach 150\%.

There are a variety of possible explanations.  We had to digitize plots 
published for sample physical cases, and either digitization errors or 
differences in the published thermal structure relative to our models could 
contribute to the discrepancies. There could also be systematic differences 
in the accelerations for the same physical conditions that should be explored. 
In our view it would be most fruitful to simply perform a direct comparison 
rather than to speculate.  Until such a direct comparison can be made, it is 
reasonable to infer that different sources for opacities can produce 
differences at the factor of 2 to 5 level in the accelerations.
We do note that the differences found for other elements were larger than the
discrepancies between the two sets of carbon accelerations, suggesting that
the relative trends (bigger differences for heavier species) are correct.

The new OP data are in better agreement with the OPAL data than was the case 
for the older OP data (Badnell \& Seaton 2003; Seaton \& Badnell 2004; 
Badnell et al. 2004), and by 
extension we expect that the accelerations will also be closer as seen in the 
case of carbon.  However, there
are real differences in the relative opacities of different species between the
OPAL data and the new OP data, and there will be significant differences in the
accelerations that we will discuss when we will have access to the OPAL data.

There are two classes of direct astrophysical implications, one for the 
interpretation of stellar surface abundance anomalies and the other for 
tests of the opacities themselves.

The differences in radiative accelerations will have important
consequences on the micro-diffusion processes. The balance between 
thermal diffusion, gravitational settling and radiative levitation will be 
significantly modified. Lower values of $g_{rad}$ could reduce the surface
abundance anomalies, making it easier to reconcile the theoretical values
with observations. It could also change the abundance profiles. Indeed the 
depth at which one species is supported is directly linked to the balance
between the 3 components of the diffusion processes. As a consequence, the Fe 
convection zone predicted in some F star models could lay deeper or even 
disappear. 
Levitation is only a perturbation in the solar case, and differences 
are unlikely to produce significant modification in the calibrated solar-like 
models. However, as the effective temperature increases, the effects will 
increase too. Therefore, theoretical errors need to be accounted for when 
comparing observations and theory.

More broadly, opacities play a crucial role in the theory of stellar 
structure and evolution. It has been difficult to establish the uncertainty 
in opacity calculations because of their complexity and the difficulty in 
obtaining direct measurements of opacity for plasmas under stellar interiors 
conditions of temperature and density.  The sensitivity of radiative 
acceleration calculations to the monochromatic opacities, however, may
provide a useful test of the opacities themselves in cases where the 
physical situation is relatively simple.  By extension, such comparisons 
will be useful for establishing the quality of opacity data across the entire 
stellar evolution field.

\acknowledgments
We would like to thank Carlos Iglesias for providing us with the carbon and 
total monochromatic opacity data from the OPAL project.  We would also like 
to thank Michael Seaton, Carlos Iglesias, and Anil Pradhan for useful 
discussions.  FD thanks Georges Alecian and Claude Zeippen for the fruitful 
discussions and collaborations during his stays in France.
Part of this work was performed during stays by FD at the Observatoire
de Paris, Meudon (France). FD is indebted to the INSU (CNRS, France) for
support during those periods.

\clearpage
\begin{figure}

\plottwo{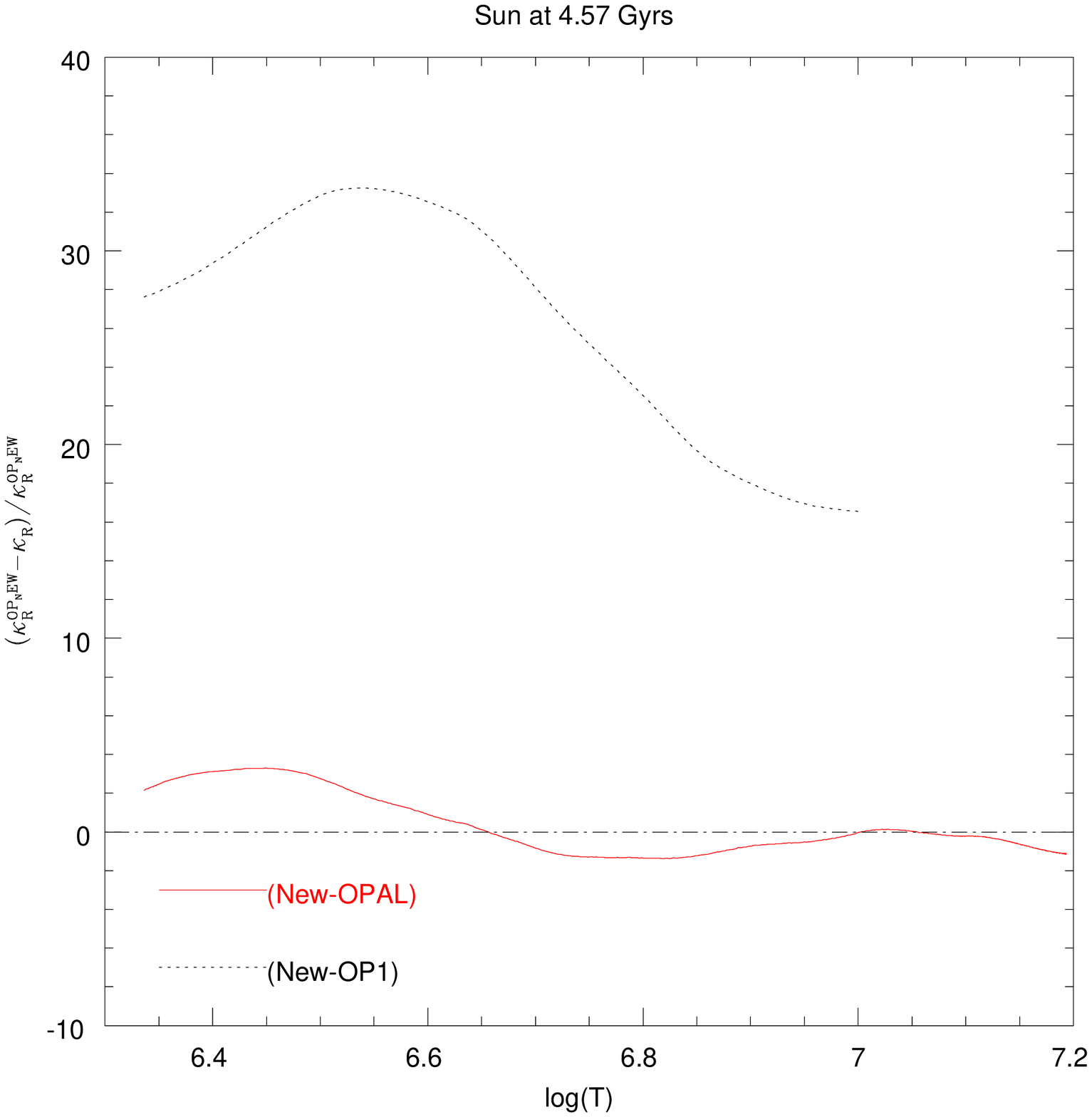}{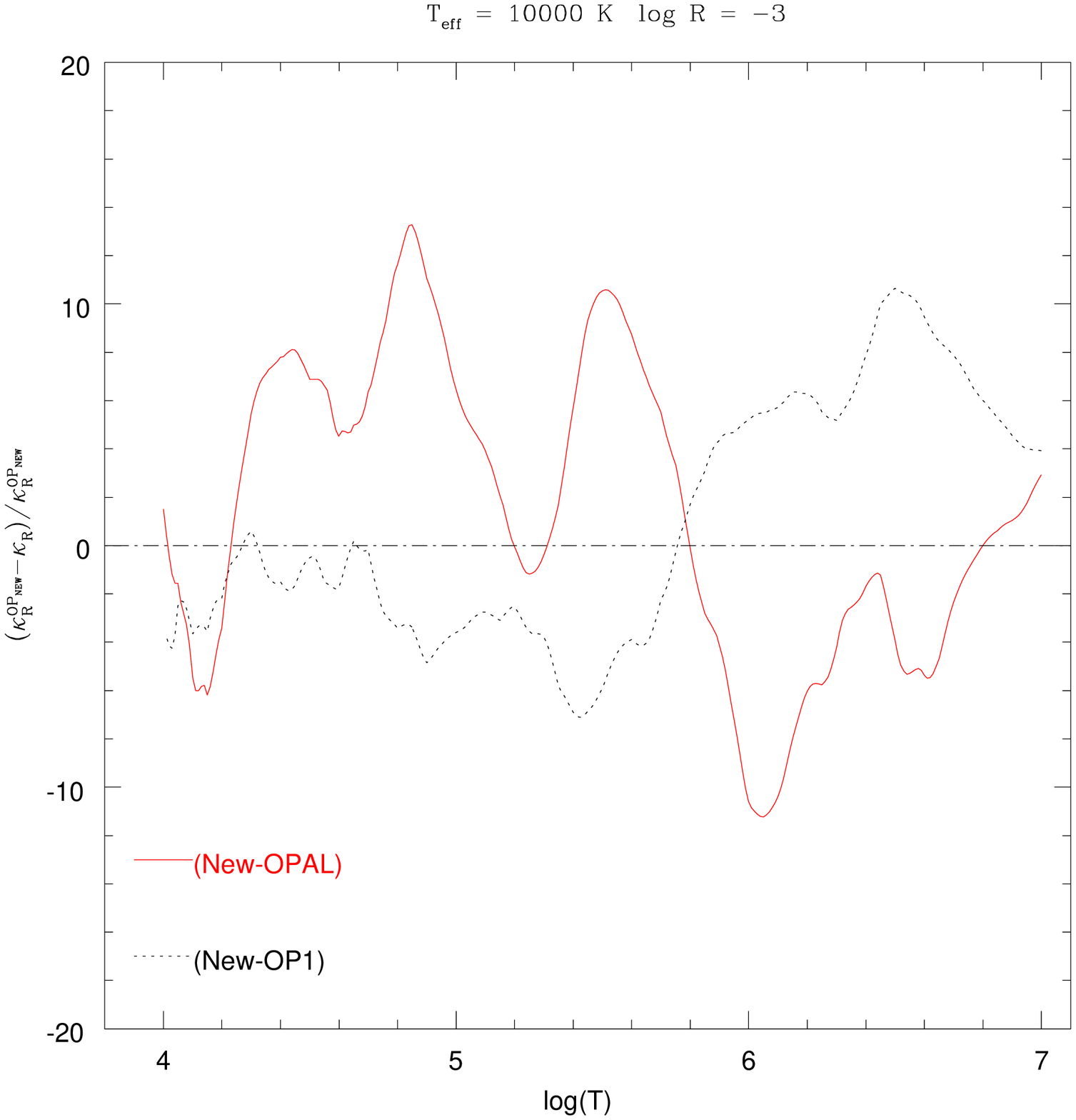}

\plottwo{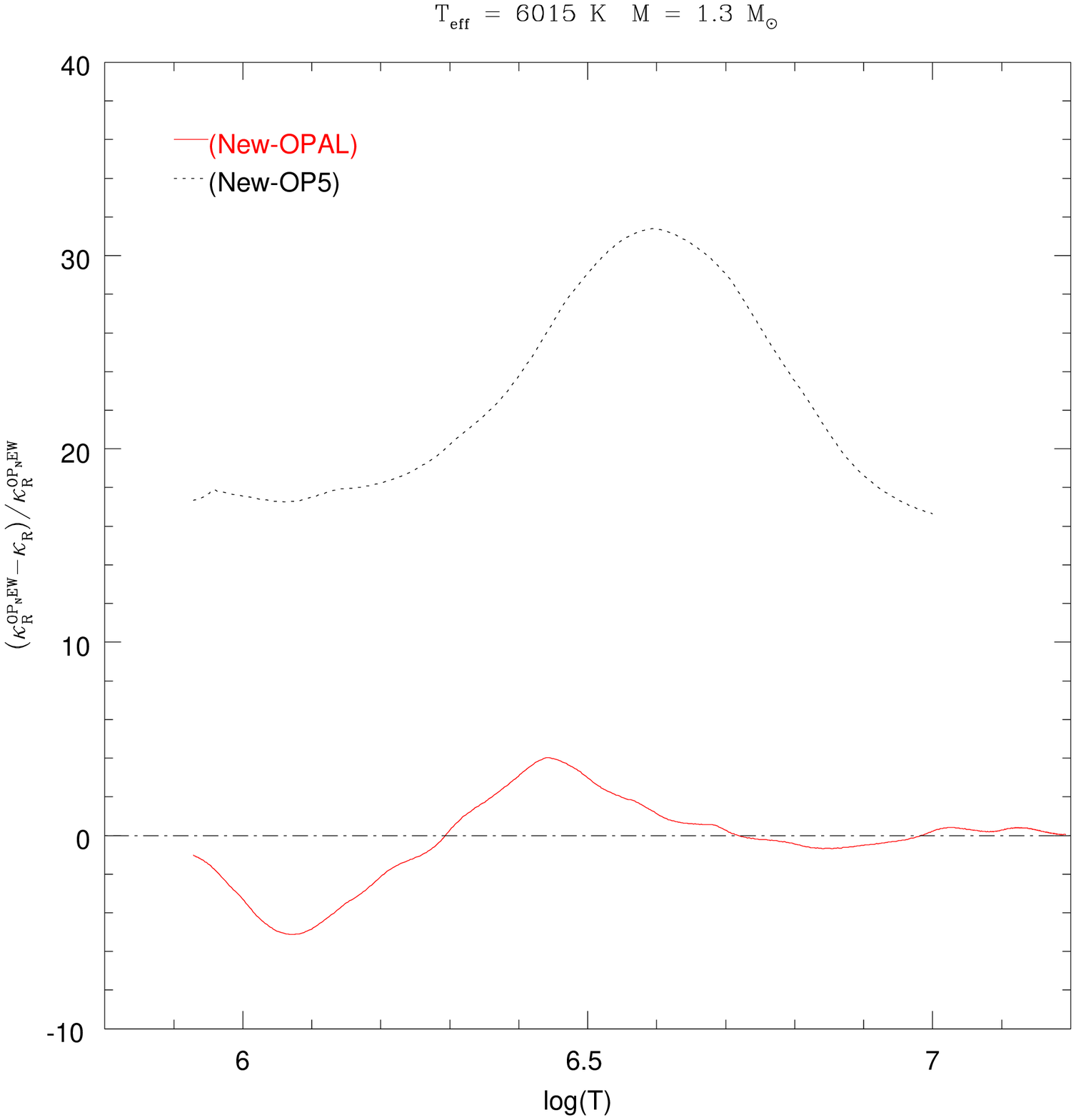}{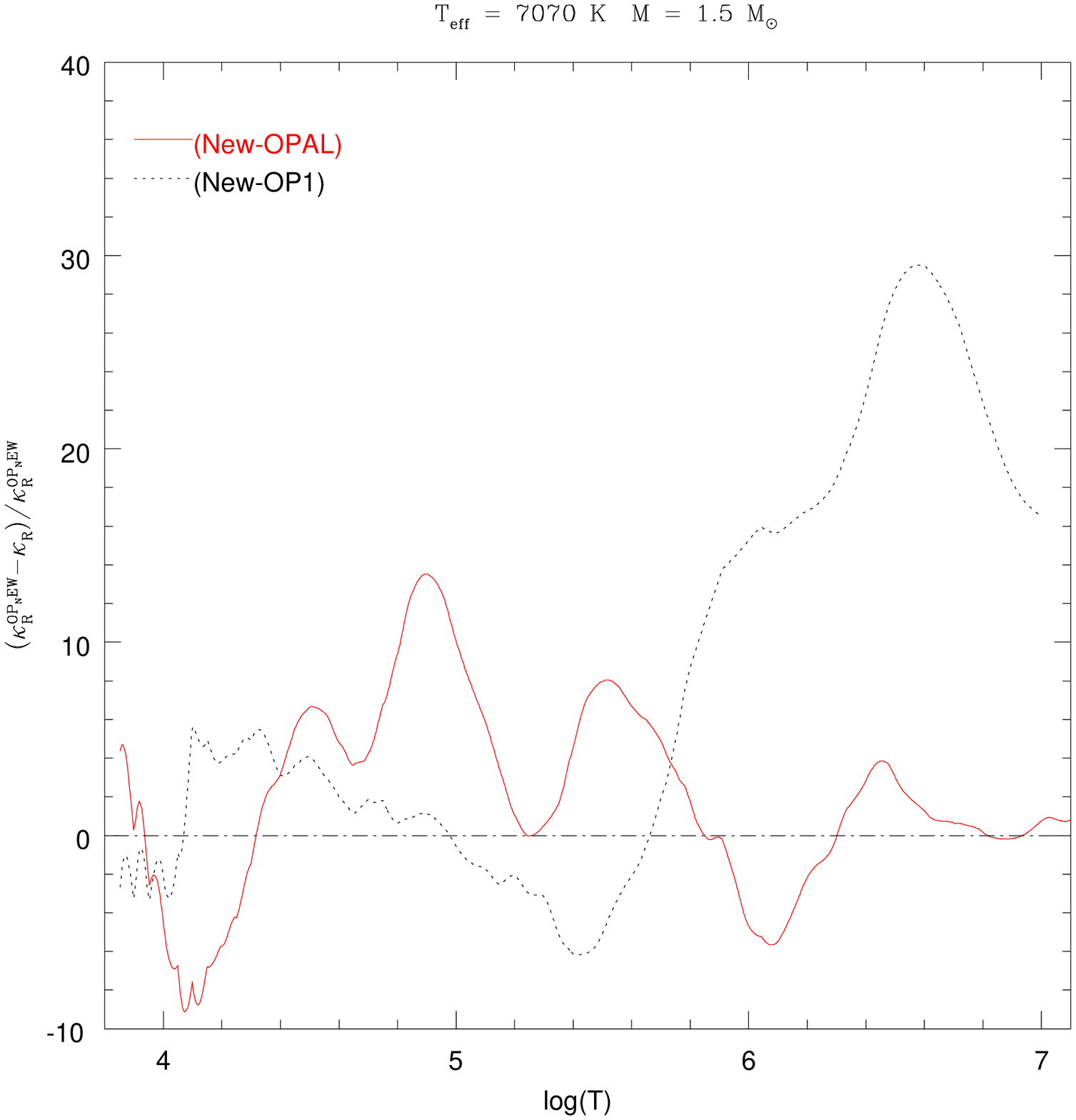}

\caption{ Percentage  difference in $\kappa_R$ between OPAL and OP:
upper left: Sun at 4.57Gyr, upper right: $T_{eff}=10000K\  Log(R)=-3$,
lower left: $M=1.3M_{\odot}\ T=6500K$, lower right: $M=1.5M_{\odot}\ T=7070K$}

\end{figure}

\begin{figure}
\plotone{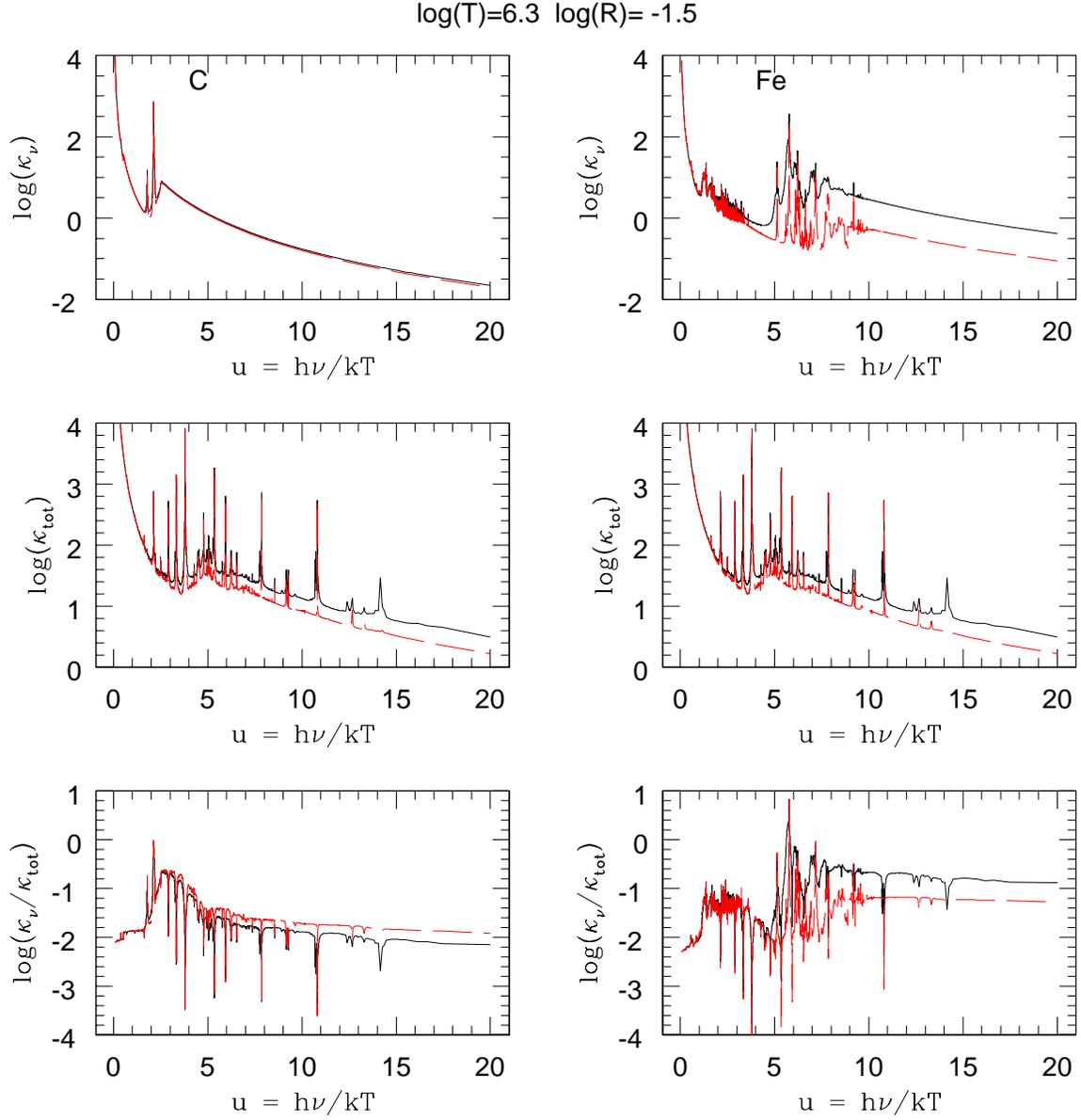}
\caption{ Monochromatic opacities at log(T)=6.3 and log(R)=-1.5: 
 Left : for Carbon, Right : for Iron. 
 Top : monochromatic opacities, Middle: total and 
 Bottom: $\frac{\kappa_{\nu}(C)}{\kappa_{\nu}^{tot}}$. 
 Dashed line $\rightarrow$ OP1 and 
 solid line $\rightarrow$ New OP}
 
\end{figure}
%

%
\begin{figure}
\plotone{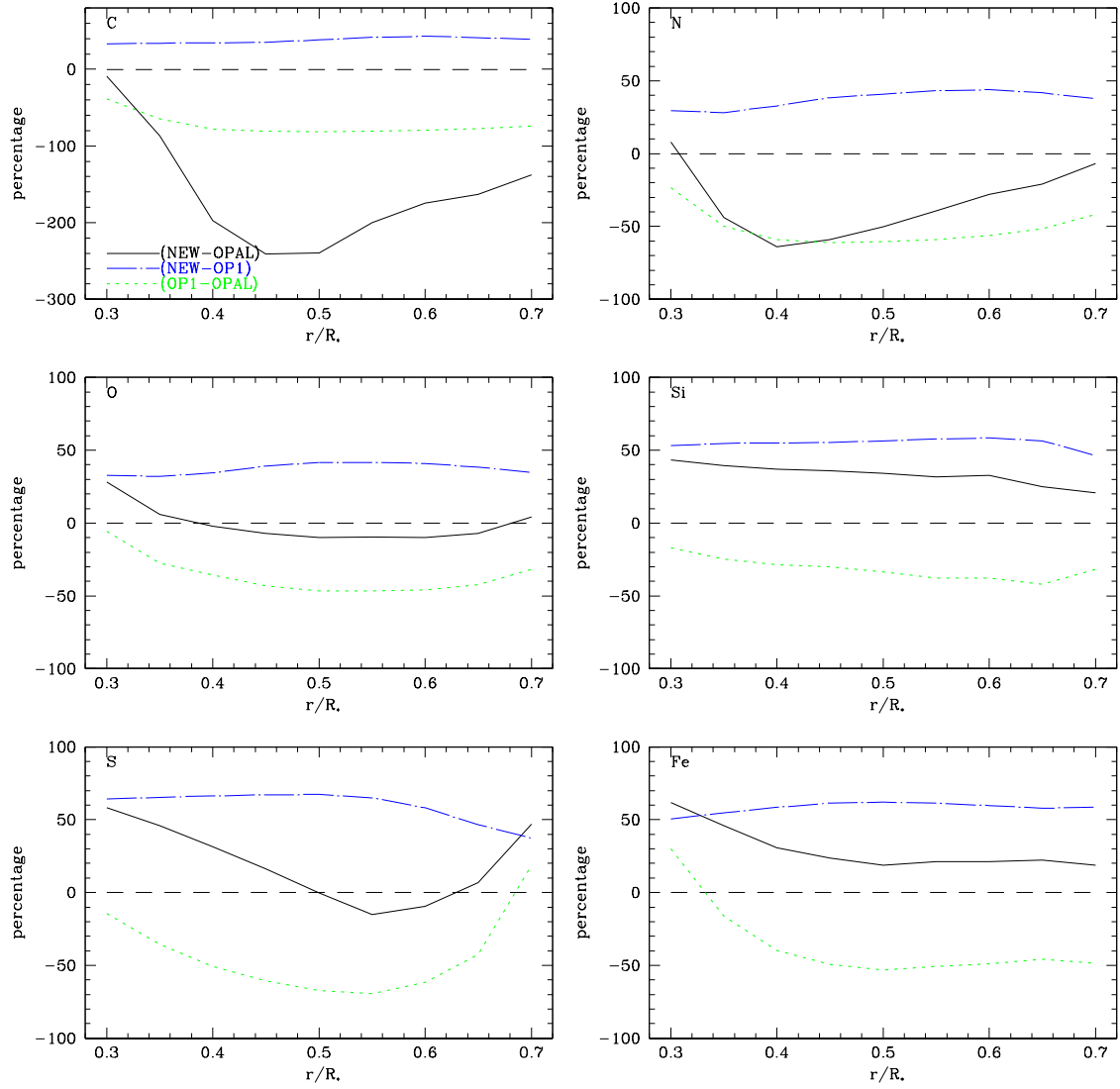}
\caption{ Percentage difference in acceleration for the solar model.}
\end{figure}
%
%
\begin{figure}
\plotone{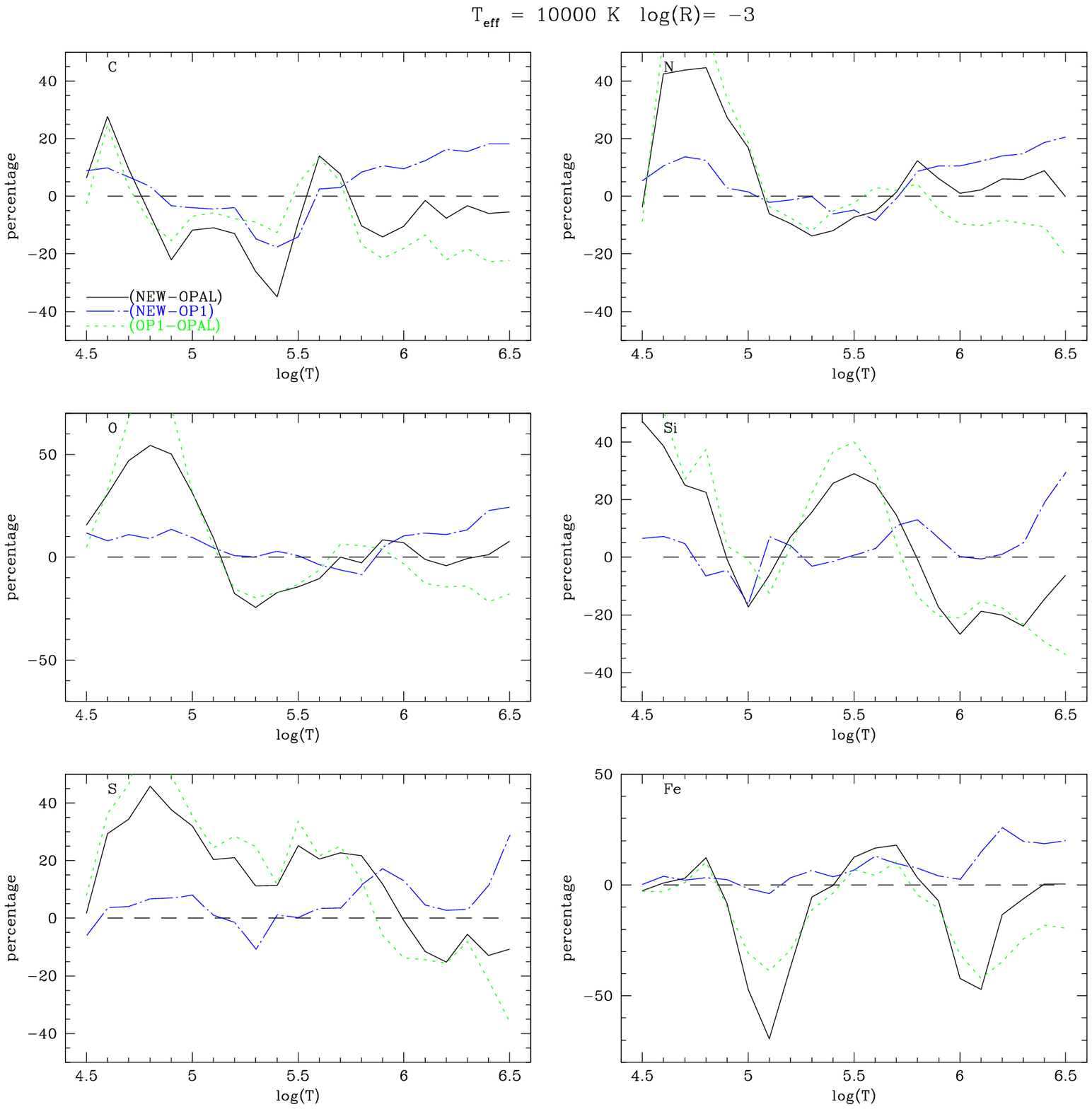}
\caption{ Percentage difference in acceleration for $T_{eff}=10000K\ logR=-3$}
\end{figure}
%
%
\begin{figure}
\plotone{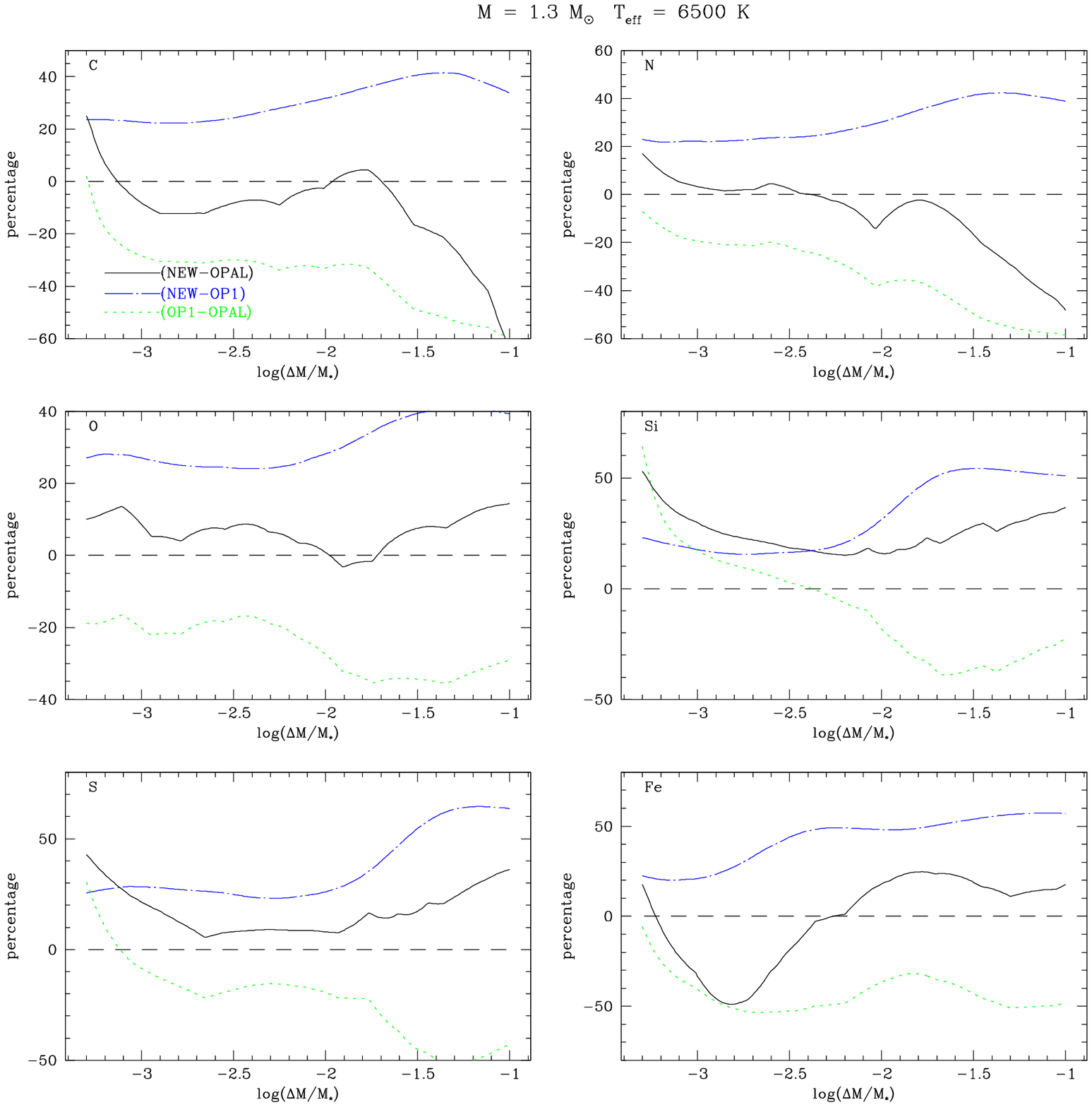}
\caption{ Percentage difference in acceleration for $M=1.3M_{\odot}$.}
\end{figure}
%
%
\begin{figure}
\plotone{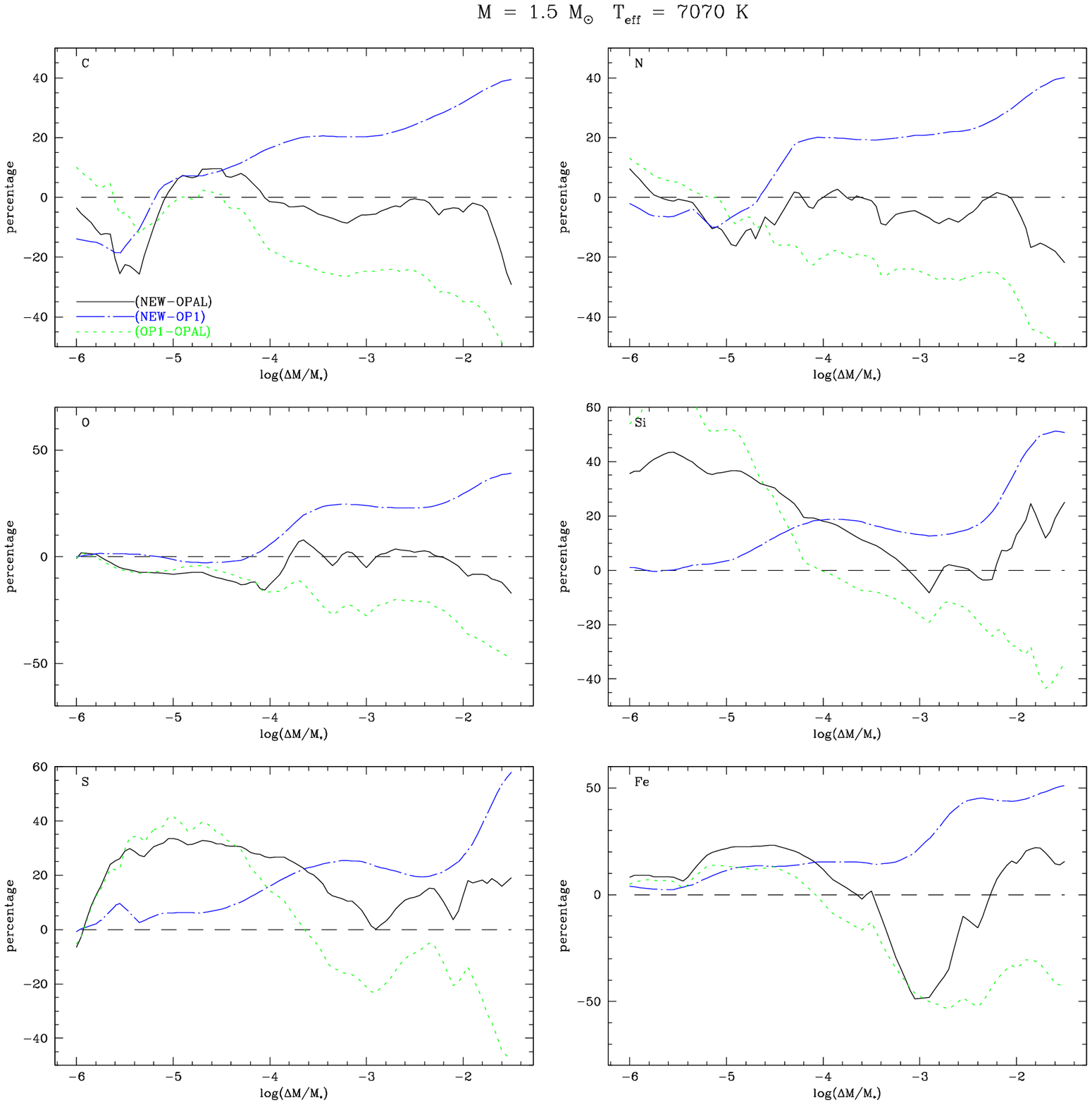}
\caption{ Percentage difference in acceleration for $M=1.5M_{\odot}$.}
\end{figure}
%
\clearpage

 \begin{table}
\centering
\begin{tabular}{cccc}
\hline
\noalign{\smallskip}
 Case & $Log(T)$ & $Log(R)$ & $Log(\rho)$  \\
\noalign{\smallskip}
            \noalign{\smallskip}
\hline
\noalign{\smallskip}
 1 & 6.0 & -1.5 & -1.5 \\
 2 & 6.0 & -2.0 & -2.0 \\
 3 & 6.0 & -2.5 & -2.5 \\
 4 & 6.0 & -3.0 & -3.0 \\
 5 & 6.3 & -1.5 & -0.6 \\
 
\noalign{\smallskip}
\hline\\
\end{tabular}
\\Table 1: ($\rho$-T) points at which individual monochromatic opacity
and total opacities have been compared.
For comparison, at the base of the solar convection zone in our model,
$Log(T)=6.33$ and $Log(\rho)=-0.735$.

\end{table}

 \begin{table}
\centering
\begin{tabular}{crc}
\hline
\noalign{\smallskip}
 Element & Z  & Number fraction \\
\noalign{\smallskip}
            \noalign{\smallskip}
\hline
\noalign{\smallskip}
 H  &   1 &   9.071E-01\\
 He &   2 &   9.135E-02\\
 C  &   6 &   3.770E-04\\
 N  &   7 &   9.913E-05\\
 O  &   8 &   7.877E-04\\
 Ne &  10 &   1.277E-04\\
 Na &  11 &   2.271E-06\\
 Mg &  12 &   4.039E-05\\
 Al &  13 &   3.135E-06\\
 Si &  14 &   3.769E-05\\
 S  &  16 &   1.722E-05\\
 Ar &  18 &   3.518E-06\\
 Ca &  20 &   2.434E-06\\
 Cr &  24 &   5.047E-07\\
 Mn &  25 &   2.608E-07\\
 Fe &  26 &   3.359E-05\\
 Ni &  28 &   1.985E-06\\
 
\noalign{\smallskip}
\hline\\
\end{tabular}
\\Table 2: Compositon .                                               

\end{table}

\clearpage
 \begin{table}
\centering
\begin{tabular}{crrr}

\hline
\noalign{\smallskip}
 Model &(New - OPAL)&(New - OP1)&( OPAL - OP1)\\

\noalign{\smallskip}
            \noalign{\smallskip}
\hline
\noalign{\smallskip}
 Sun & 1.3\% & 29\% &  29\% \\
 10000 & 6.2\% & 6.5\% &  9.5\% \\
 $M=1.3M_{\odot}$ & 1.9\% & 27\% &  26.9\% \\
 $M=1.5M_{\odot}$ & 4.8\% &  17.4\%&   17.6\% \\
 
\noalign{\smallskip}
\hline\\
\end{tabular}
\\Table 3: rms of the percentage difference for all points of each model.
$rms = \sqrt{\frac{\sum (\frac{\Delta \kappa_R}{\kappa_R})^2}{N}}$

\end{table}

 \begin{table}
\centering
\begin{tabular}{ccrrrcc}

\hline
\noalign{\smallskip}
 log(T) & $log(\rho)$ & $\kappa^{Present}_R$ & $\kappa^{OPAL}_R$ & 
  $\kappa^{OP5}_R$ & (NEW-OPAL)& (NEW-OP1)\\

\noalign{\smallskip}
            \noalign{\smallskip}
\hline
\noalign{\smallskip}
 6.0& -1.5 & 58.75 &  60.38 (60.37) & 44.63 & -2.7\%  & 24.1\% \\
 6.0& -2.0 & 27.54 &  28.83 (28.90) & 22.10 & -4.7\%  & 19.8\%\\
 6.0& -2.5 & 10.29 &  11.42 (11.52) &  8.93 & -10.9\%  & 13.2\%\\
 6.0& -3.0 &  3.41 &   3.65 ( 3.72) &  3.22 & -7.0\%  &  5.6\%\\
 6.3& -0.6 & 34.89 &  35.79 (35.76) & 24.61 & -2.6\%  & 29.5\%\\
 
\noalign{\smallskip}
\hline\\
\end{tabular}
\\Table 4:  OPAL and OP $\kappa_R$ and $\gamma_C$.
The percentage difference correspond to
$\frac{\kappa^{OP}_R-\kappa^{OPAL}_R}{\kappa^{OPAL}_R}$.
 The OPAL Rosseland mean has
been recalculated using a sampling/interpolation procedure in order to
have the same frequency point. This procedure has an error of less than 1\%
except for point 4 ($log(T)=6.0$ $log(\rho)=-3.0$) for which the difference 
between the recalculated value (in parenthesis) and the value given by OPAL 
is 1.8\%.

\end{table}

\clearpage
 \begin{table}
\centering
\begin{tabular}{ccccrrr}
\hline
\noalign{\smallskip}
 logT & $log(\rho)$ &$log(T_{eff})$&$r/R_{*}$& $log(\gamma^{New})$ &
 $log(\gamma^{OPAL})$ & 
 $\frac{\delta \gamma_1}{\gamma}$ \\
\noalign{\smallskip}
            \noalign{\smallskip}
\hline
\noalign{\smallskip}
 6.0& -1.5 &3.762&0.7146 &2.668 & 2.664 & 0.9 \% \\ 
 6.0& -2.0 &3.762&0.7146 &2.583 & 2.579 & 0.9 \%\\ 
 6.0& -2.5 &3.762&0.7146 &2.491 & 2.476 & 3.4 \% \\ 
 6.0& -3.0 &3.762&0.7146 &2.406 & 2.407 & -0.2 \%\\ 
 6.3& -0.6 &3.762&0.7146 &2.239 & 2.188 & 11.1 \%\\

\noalign{\smallskip}
\hline\\
\end{tabular}
\\ Table 5: Comparison of $\gamma(C)$ obtained with OPAL and OP data 
 (New). The effects of momentum transfer to the electron are not taken into 
 account.\\ 
Note: $\frac{\delta \gamma_1}{\gamma} = \frac{\gamma^{New}-\gamma^{OPAL}}{\gamma^{New}}$; 

\end{table}

 \begin{table}
\centering
\begin{tabular}{ccccrrcrrc}
\hline
\noalign{\smallskip}
 logT & $log(\rho)$ &$log(T_{eff})$&$r/R_{*}$& $log(\gamma^{New})$ &
 $log(\gamma^{OPAL})$ & $log(\gamma^{OP1})$&
 $\frac{\delta \gamma_1}{\gamma}$ & $\frac{\delta \gamma_2}{\gamma}$ & 
 $\frac{\delta \gamma_3}{\gamma}$\\
\noalign{\smallskip}
            \noalign{\smallskip}
\hline
\noalign{\smallskip}
  6.0& -1.5 &3.762&0.7146&2.621 & 2.625 & 2.585 & -0.9\% &  8.0\% & -8.8\%\\ 
  6.0& -2.0 &3.762&0.7146&2.526 & 2.535 & 2.496 & -2.1\% &  6.7\% & -8.6\%\\ 
  6.0& -2.5 &3.762&0.7146&2.420 & 2.424 & 2.396 & -0.9\% &  5.4\% & -6.2\%\\ 
  6.0& -3.0 &3.762&0.7146&2.322 & 2.343 & 2.299 & -5.0\% &  5.2\% & -9.6\%\\ 
  6.3& -0.6 &3.762&0.7146&2.089 & 2.060 & 2.020 & 6.5\% &  14.7\% & -8.8\%\\

\noalign{\smallskip}
\hline\\
\end{tabular}
\\ Table 6: Comparison of $\gamma(C)$ obtained with OPAL and OP data 
 (New and OP5). The effect of momentum tansfer to the electron is taken onto account.\\ 
Note: $\frac{\delta \gamma_1}{\gamma} = \frac{\gamma^{New}-\gamma^{OPAL}}{\gamma^{New}}$; 
    $\frac{\delta \gamma_2}{\gamma} = \frac{\gamma^{New}-\gamma^{OP5}}{\gamma^{New}}$; 
    $\frac{\delta \gamma_3}{\gamma} = \frac{\gamma^{OP}-\gamma^{OPAL}}{\gamma^{OPAL}}$

\end{table}

\clearpage
 \begin{table}
\centering
\begin{tabular}{ccccrrr}
\hline
\noalign{\smallskip}
 logT & $log(\rho)$ &$log(T_{eff})$&$r/R_{*}$& $log(g_{rad}^{New})$ &
 $log(g_{rad}^{OPAL})$ & $\frac{\delta g_1}{g}$ \\
\noalign{\smallskip}
            \noalign{\smallskip}
\hline
\noalign{\smallskip}
  6.0& -1.5 &3.762&0.7146&4.077 & 4.097 & -4.7\% \\
  6.0& -2.0 &3.762&0.7146&3.664 & 3.692 & -6.7\% \\
  6.0& -2.5 &3.762&0.7146&3.145 & 3.190 & -10.9\% \\
  6.0& -3.0 &3.762&0.7146&2.580 & 2.629 & -11.9\% \\
  6.3& -0.6 &3.762&0.7146&3.423 & 3.393 & 6.7\% \\

\noalign{\smallskip}
\hline\\
\end{tabular}
\\ Table 7: Comparison of the C acceleration obtained with OPAL and OP data 
 (New). The accelerations do not take into account the effect of
  momentum tansfer to the electron. When the correction is applied, the 
  the differences increase.\\
   Note: $\frac{\delta g_1}{g} = \frac{g^{New}-g^{OPAL}}{g^{New}}$; 

\end{table}

 \begin{table}
\centering
\begin{tabular}{ccccrrcrrc}
\hline
\noalign{\smallskip}
 logT & $log(\rho)$ &$log(T_{eff})$&$r/R_{*}$& $log(g_{rad}^{New})$ &
 $log(g_{rad}^{OPAL})$ & $log(g_{rad}^{OP1})$&
 $\frac{\delta g_1}{g}$ & $\frac{\delta g_2}{g}$ & $\frac{\delta g_3}{g}$\\
\noalign{\smallskip}
            \noalign{\smallskip}
\hline
\noalign{\smallskip}
  6.0& -1.5 &3.762&0.7146&4.030 & 4.058 & 3.887 & -7\% & 28\% & -32\%\\ 
  6.0& -2.0 &3.762&0.7146&3.607 & 3.648 & 3.494 & -10\% & 23\% & -30\%\\ 
  6.0& -2.5 &3.762&0.7146&3.074 & 3.138 & 3.000 & -16\% & 16\% & -27\%\\ 
  6.0& -3.0 &3.762&0.7146&2.495 & 2.565 & 2.460 & -17\% & 8\% & -21\%\\ 
  6.3& -0.6 &3.762&0.7146&3.273 & 3.266 & 3.064 & 2\% & 38\% & -37\%\\

\noalign{\smallskip}
\hline\\
\end{tabular}
\\ Table 8: Same as table 7 but the momentum transfer to the electron
   is taken into account as well as the $e^-$ scatering has been removed from
   the carbon mocrochromatic opacities using the OP data. 
   Note: $\frac{\delta g_1}{g} = \frac{g^{New}-g^{OPAL}}{g^{New}}$; 
    $\frac{\delta g_2}{g} = \frac{g^{New}-g^{OP1}}{g^{New}}$; 
    $\frac{\delta g_3}{g} = \frac{g^{OP1}-g^{OPAL}}{g^{OPAL}}$

\end{table}

\enddocument